\documentclass[usenatbib,usegraphicx]{mn2e}
\usepackage{amssymb}

\DeclareGraphicsExtensions{.eps,.ps}

\newcommand{\upf}[1]{\raisebox{0.5ex}[0pt]{#1}}

\title[Star clusters in the Sh2-132 complex]{Star clusters in the Sh2-132 complex:
clues about the connection between embedded and open clusters}

\author[T.A. Saurin, E. Bica and C. Bonatto]{T.A. Saurin$^{1}$\thanks{E-mail:
tiago.saurin@ufrgs.br (TAS)}, E. Bica$^{1}$\thanks{E-mail: bica@if.ufrgs.br (EB)}
and C. Bonatto$^{1}$\thanks{E-mail: charles@if.ufrgs.br (CB)}\\
$^{1}$Universidade Federal do Rio Grande do Sul, Departamento de Astronomia,
CP\,15051, RS, Porto Alegre 91501-970, Brazil}

\begin{document}

\date{Accepted --. Received --}

\pagerange{\pageref{firstpage}--\pageref{lastpage}} \pubyear{2009}

\maketitle

\label{firstpage}

\begin{abstract}
Embedded clusters are formed in molecular clouds where massive stars can
produce HII regions. The detailed embedded-open cluster evolutionary
connection as well as the origin of associations are yet to be unveiled.
There appears to be a high infant mortality rate among embedded clusters
and the few survivors evolve to open clusters.
We study the colour-magnitude diagrams and structure of the star
clusters related to the Sh2-132 HII region using the 2MASS database.
Cluster fundamental and structural parameters are determined via MS and PMS
isochrones and stellar radial density profiles.
We report the discovery of four clusters. One of them is projected a few
diameters away from the optical cluster Teutsch\,127 and appears to be
deeply embedded, seen only in the infrared. Evidence is found that we are
witnessing the dynamical transition from an embedded to an open cluster.
An additional cluster is also close to Teutsch\,127 and might be
associated with a bow-shock. We also study the CMD and structure of the
open cluster Berkeley\,94 in Sh2-132 and a new cluster which is projected in
the outskirts of the complex. Finally, we searched for star clusters
around the two known Wolf-Rayet stars in the complex. One of them appears to
be related to a compact cluster.
Finally, the present analyses suggest early dynamical evolution
for young star clusters.
\end{abstract}

\begin{keywords}
stars: formation --
open clusters and associations: individual: Teutsch\,127 --
open clusters and associations: individual: Berkeley\,94
\end{keywords}

\section{Introduction}
\label{sec:int}
Early star cluster evolution generally leads to dissolution into the field
due to rapid gas expulsion associated with evolutionary effects of massive stars
such as supernovae and stellar winds. This causes abrupt changes in the
gravitational potential, to the point that significant fractions of the low-mass
stars with high velocity in the outer region escape from the cluster (e.g.
\citealt{tutu78}, \citealt{felkro05}, \citealt{basgie08}). Bound open clusters
(OCs), i.e. those that keep dynamical stability after the first
$\sim$\,30--40\,Myr, appear to be minority of about 5\%
(\citealt{ladlad03}). There is a high discrepancy between the birthrate of
embedded clusters (2--4\,Myr$^{-1}$\,kpc$^{-2}$, \citealt{ladlad03}) and OCs
(0.25\,Myr$^{-1}$\,kpc$^{-2}$ according to \citealt{elcl85}, and
0.45\,Myr$^{-1}$\,kpc$^{-2}$ according to \citealt{baca91}) within 2\,kpc of the
Sun. This is likely a consequence of the high rate of infant mortality.
The cluster survival depends primarily on the dynamical state
of the stars, i.e., their virial ratio immediately before the onset of gas
expulsion \citep{good09}, while the star-formation efficiency has a
secondary role. As an example of a dissolving cluster, a recent study showed
that Bochum\,1 has an eroded radial density profile (RDP) and a detached core,
suggesting, in that case, the genesis of a small OB association
(\citealt*{bicbon08}).

\begin{figure*}
\resizebox{\hsize}{!}{\includegraphics{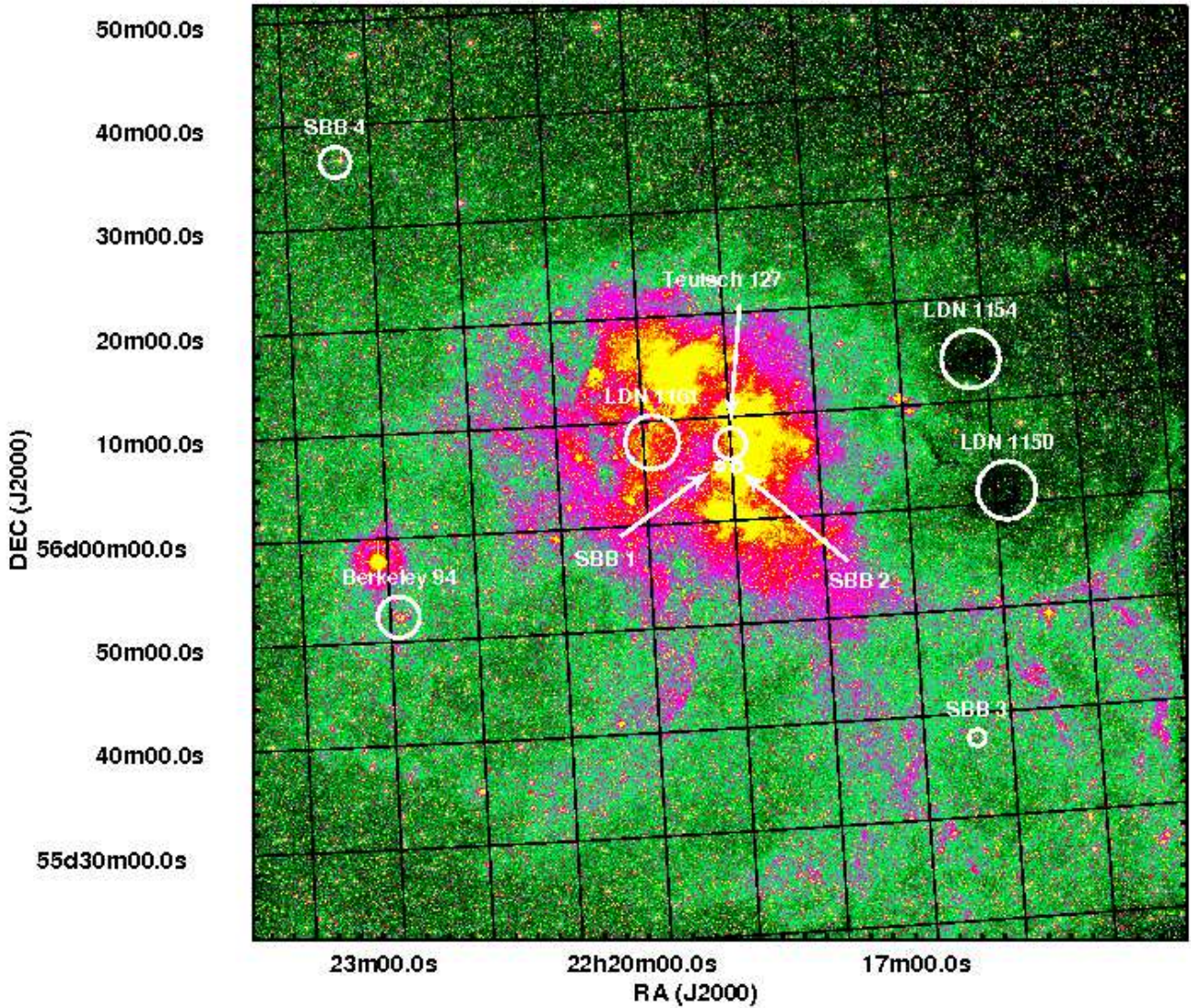}}
\caption[]{XDSS $R$-band image in false colours of Sh2-132 with all star clusters
of the present study (circle radii are listed in Table\,\ref{tab:clu}).
Dark nebulae in the area are indicated by circle radii 2.83$'$ for
LDN\,1150 and LDN\,1154, and 2.62$'$ for LDN\,1161 (\citealt{lynds62}). A large
bubble appears to be related to the cluster Berkeley\,94, and another one
encloses the dark nebulae LDN\,1150 and LDN\,1154}
\label{fig:sh132}
\end{figure*}

Young OCs are probes of the early physical processes occuring in embedded and
post-embedded clusters. In recent years our group has analyzed main sequence
(MS) and pre-main sequence (PMS) stellar contents of such objects using powerful
tools for field decontamination (NGC\,4755 in \citealt{bbob06}; NGC\,6611 in
\citealt*{bsb06}; Pismis\,5, NGC\,1931, vdB\,80 and BDBS\,96 in
\citealt{bonbic09a}; and NGC\,2244 in \citealt{bonbic09b}).

The development of infrared astronomy during the last years has improved
our understanding about the origin of clusters, but there is a lot of questions
to be answered yet. Can all intermediate steps of early cluster evolution be
tracked? Can evolutionary scenarios be tested? The present work together with
other recent ones by our group address this puzzle, which is fundamental for the
understanding of star cluster and field populations build up in the galaxy.

Another fundamental question is whether massive stars can be formed in
isolation, or if isolated massive stars are remnants of dispersed embedded
clusters. \citet{wit04} and \citet{wit05} found that 4\,$\pm$\,2\% of all O-type
stars with $V$\,$<$\,8\,mag could have been formed outside a cluster.
\citet{pargoo07} have suggested that some apparently isolated O-stars are in
fact low-mass clusters ($<$\,100\,M$_{\odot}$) with only one massive star.
N-body simulations by \citet{pfkro06} showed that the Orion Nebula cluster could
have lost at least half of its OB stars as runaways during early cluster
evolution. The presence of bow-shocks associated with massive stars can be used
as an indicator of ejection of these stars at supersonic velocities
(e.g. \citealt{gvabom09}), when the proper motion cannot be measured properly
because of large distances and shortly separated epochs. However, only a
fraction $\lesssim$\,20\% of runaway OB stars produce bow-shocks.

In the present study, we explore a star-forming complex related to the Sh2-132
HII region (\citealt{sharpl59}) that embeds some previously undetected clusters.
Using 2MASS\footnote{\it http://www.ipac.caltech.edu/2mass} photometry we
analyse colour-magnitude diagrams (CMDs) and RDPs of 5 star clusters related or
possibly related to Sh2-132 HII. An additional star cluster projected close to
the complex is also studied. A radio continuum emission analysis of Sh2-132
(\citealt*{hafeto78}) indicated the giant nature of this complex and its evolved
character (low electron densities, large linear diameter and an ionized gas
mass greater than 10$^4$\,M$_{\odot}$), which suggests a suitable laboratory
for early dynamical and hydrodynamical evolution.

We also investigate two Wolf-Rayet (WR) stars in the area, searching for
possible star clusters around them.

In Sect.\,\ref{sec:dir} we present the targets. In Sect.\,\ref{sec:cmd} we carry
out CMD analysis and estimate cluster parameters. In Sect.\,\ref{sec:rdp} we
study the density profiles. Discussions are given in Sect.\,\ref{sec:dis}.
Finally, conclusions of this work are in Sect.\,\ref{sec:con}.

\section{Newly found and previously known star clusters}
\label{sec:dir}
The number of embedded star clusters in a given star-forming region is
essential to study the gas expulsion and dynamical evolution effects, in order
to know about cluster binarity or multiplicity (\citealt{carva08}). In this
investigation line we searched for new embedded clusters in Sh2-132 complex.

We present in Table\,\ref{tab:clu} the star clusters in the area of the Sh2-132.
The radii given in Col.\,5 were chosen by eye for decontamination purposes
(Sect.\,\ref{sec:cmd}), since they provide a good contrast with the field.
Fig.\,\ref{fig:sh132} shows an $R$-band 2nd generation Digitized Sky Survey
(XDSS\footnote{\it http://www1.cadc-ccda.hia-iha.nrc-cnrc.gc.ca/cadc})
image of the whole complex with the positions of the star clusters
(Table\,\ref{tab:clu}) together with the dark nebulae LDN\,1150, LDN\,1154 and
LDN\,1161 (\citealt{lynds62}). SBB\,1, SBB\,2, SBB\,3 and SBB\,4
were found by one of us (TAS) using the 2MASS Atlas. The optical OC
Teutsch\,127 (\citealt*{kretal06}) contains a trapezium system (Trap\,900) found
by \citet{abtcor00}. Berkeley\,94 was studied with photographic photometry by
\citet{yilmaz70} and the derived parameters are given in
WEBDA\footnote{\it http://www.univie.ac.at/webda/webda.html}: age
$t$\,=\,10\,Myr, reddening {\it E(B$-$V)}\,=\,0.61 and distance from the Sun
{\it d$_{\odot}$}\,=\,2.63\,kpc. It is important to analyse Berkeley\,94 also
with 2MASS photometry for comparison purposes.

\begin{table*}
\caption[]{Known and newly found star clusters in the Sh2-132 area.}
\label{tab:clu}
\renewcommand{\tabcolsep}{2.8mm}
\renewcommand{\arraystretch}{1.25}
\begin{tabular}{ccccccl}
\hline
\multicolumn{1}{c}{$l$} &\multicolumn{1}{c}{$b$} &\multicolumn{1}{c}{$\alpha$}
&\multicolumn{1}{c}{$\delta$} &\multicolumn{1}{c}{$r$} &\multicolumn{1}{c}{Name}
&\multicolumn{1}{c}{Comments} \\
\multicolumn{1}{c}{(1)} &\multicolumn{1}{c}{(2)} &\multicolumn{1}{c}{(3)}
&\multicolumn{1}{c}{(4)} &\multicolumn{1}{c}{(5)} &\multicolumn{1}{c}{(6)}
&\multicolumn{1}{c}{(7)} \\
\hline
102$^{\circ}_{\upf{.}}$81 & -0$^{\circ}_{\upf{.}}$67 & 22$^{\rm h}$19$^{\rm m}$00$^{\rm s}_{\upf{.}}$0 & 56$^{\circ}$07$'$25$''_{\upf{.}}$0 & 1$'_{\upf{.}}$5 & Teutsch\,127$^1$ & optical embedded open cluster, includes Trap\,900 \\
102$^{\circ}_{\upf{.}}$81 & -0$^{\circ}_{\upf{.}}$71 & 22$^{\rm h}$19$^{\rm m}$08$^{\rm s}_{\upf{.}}$2 & 56$^{\circ}$05$'$17$''_{\upf{.}}$5 & 0$'_{\upf{.}}$4 & SBB\,1 & infrared embedded cluster with IRAS\,22172+5549 \\
102$^{\circ}_{\upf{.}}$78 & -0$^{\circ}_{\upf{.}}$70 & 22$^{\rm h}$18$^{\rm m}$56$^{\rm s}_{\upf{.}}$7 & 56$^{\circ}$05$'$10$''_{\upf{.}}$9 & 0$'_{\upf{.}}$5 & SBB\,2 & optical embedded cluster with bow-shock \\
103$^{\circ}_{\upf{.}}$13 & -1$^{\circ}_{\upf{.}}$18 & 22$^{\rm h}$22$^{\rm m}$54$^{\rm s}_{\upf{.}}$3 & 55$^{\circ}$52$'$24$''_{\upf{.}}$0 & 2$'_{\upf{.}}$0 & Berkeley\,94$^1$ & optical open cluster in the outskirts of the complex \\
102$^{\circ}_{\upf{.}}$23 & -0$^{\circ}_{\upf{.}}$86 & 22$^{\rm h}$16$^{\rm m}$24$^{\rm s}_{\upf{.}}$1 & 55$^{\circ}$37$'$37$''_{\upf{.}}$0 & 0$'_{\upf{.}}$8 & SBB\,3 & compact optical open cluster surrounding WR\,152 \\
103$^{\circ}_{\upf{.}}$58 & -0$^{\circ}_{\upf{.}}$60 & 22$^{\rm h}$23$^{\rm m}$24$^{\rm s}_{\upf{.}}$3 & 56$^{\circ}$36$'$27$''_{\upf{.}}$0 & 1$'_{\upf{.}}$5 & SBB\,4 & optical open cluster in the outskirts of the complex \\
\hline
\end{tabular}
\begin{list}{Table Notes.}
\item Col.\,1: Galactic longitude, Col.\,2: Galactic latitude,
Col.\,3: right ascencion (J2000), Col.\,4: declination (J2000),
Col.\,5: radius for decontamination, Col.\,6: star cluster name,
Col.\,7: comments. $^1$From the SIMBAD database.
\end{list}
\end{table*}

The Sh2-132 complex includes two WR stars (\citealt{vdhuch01} -- see also the
Vizier\footnote{\it http://vizier.u-strasbg.fr/viz-bin/VizieR?-source=III/215}
catalogue III/215). We list in Table\,\ref{tab:wrs} their basic properties. They
were indicated as members of the Sh2-132 complex and the OB association
Cep\,OB1. \citet{caetal08} analysed the distribution of interstellar matter in
the neighbourhood of WR\,152 and WR\,153ab with radio and optical data. They
found an HI bubble surrounding the ionized shell of WR\,152 and the
occurrence of photodissociation at the interface between the ionized and
molecular gas around WR\,153ab.

\begin{table*}
\caption[]{WR stars in Sh2-132 (\citealt{vdhuch01}).}
\label{tab:wrs}
\renewcommand{\tabcolsep}{3.6mm}
\renewcommand{\arraystretch}{1.25}
\begin{tabular}{lccccccc}
\hline
\multicolumn{1}{c}{Name} &\multicolumn{1}{c}{$\alpha$}
&\multicolumn{1}{c}{$\delta$} &\multicolumn{1}{c}{$A_V$}
&\multicolumn{1}{c}{$d$\,(kpc)} &\multicolumn{1}{c}{ST}
&\multicolumn{1}{c}{Association} &\multicolumn{1}{c}{HII region} \\
\multicolumn{1}{c}{(1)} &\multicolumn{1}{c}{(2)} &\multicolumn{1}{c}{(3)}
&\multicolumn{1}{c}{(4)} &\multicolumn{1}{c}{(5)} &\multicolumn{1}{c}{(6)}
&\multicolumn{1}{c}{(7)} &\multicolumn{1}{c}{(8)}\\
\hline
WR\,152, HD\,211564   & 22$^{\rm h}$16$^{\rm m}$24$^{\rm s}_{\upf{.}}$1 & 55$^{\circ}$37$''$37$'$ & 1.62 & 2.75 & WN3(h)-w & Cep\,OB1 & Sh2-132 \\
WR\,153ab, HD\,211853 & 22$^{\rm h}$18$^{\rm m}$45$^{\rm s}_{\upf{.}}$6 & 56$^{\circ}$07$''$37$'$ & 2.28 & 2.75 & WN6o+O6I & Cep\,OB1 & Sh2-132 \\
\hline
\end{tabular}
\begin{list}{Table Notes.}
\item Col.\,1: Wolf-Rayet star name, Col.\,2: right ascencion (J2000),
Col.\,3: declination (J2000), Col.\,4: $V$-band absorption,
Col.\,5: distance from the Sun,
Col.\,6: spectral type (\citealt{hgl06}; \citealt{ssm96}),
Col.\,7: association, Col.\,8: HII region.
\end{list}
\end{table*}

\begin{table*}
\caption[]{Early type stars in the area of the clusters (SIMBAD).}
\label{tab:ear}
\renewcommand{\tabcolsep}{7.6mm}
\renewcommand{\arraystretch}{1.25}
\begin{tabular}{lcccrc}
\hline
\multicolumn{1}{c}{Name} &\multicolumn{1}{c}{$\alpha$}
&\multicolumn{1}{c}{$\delta$} &\multicolumn{1}{c}{ST} &\multicolumn{1}{c}{$V$}
&\multicolumn{1}{c}{Related Cluster} \\
\multicolumn{1}{c}{(1)} &\multicolumn{1}{c}{(2)} &\multicolumn{1}{c}{(3)}
&\multicolumn{1}{c}{(4)} &\multicolumn{1}{c}{(5)} &\multicolumn{1}{c}{(6)} \\
\hline
BD+55\,2722$^1$	 & 22$^{\rm h}$18$^{\rm m}$59$^{\rm s}_{\upf{.}}$1 & 56$^{\circ}$07$''$23$'$ & O9V &  9.54 & Teutsch\,127 \\
TYC\,3986-3487-1 & 22$^{\rm h}$18$^{\rm m}$59$^{\rm s}_{\upf{.}}$9 & 56$^{\circ}$07$''$19$'$ & B   & 10.60 & Teutsch\,127 \\
BD+55\,2736	 & 22$^{\rm h}$22$^{\rm m}$52$^{\rm s}_{\upf{.}}$7 & 55$^{\circ}$52$''$13$'$ & B   &  9.67 & Berkeley\,94 \\
TYC\,3986-545-1  & 22$^{\rm h}$22$^{\rm m}$51$^{\rm s}_{\upf{.}}$7 & 55$^{\circ}$52$''$31$'$ & B   & 11.59 & Berkeley\,94 \\
TYC\,3986-1890-1 & 22$^{\rm h}$22$^{\rm m}$57$^{\rm s}_{\upf{.}}$6 & 55$^{\circ}$52$''$51$'$ & B   & 10.22 & Berkeley\,94 \\
TYC\,3986-695-1  & 22$^{\rm h}$22$^{\rm m}$47$^{\rm s}_{\upf{.}}$8 & 55$^{\circ}$52$''$26$'$ & B5  & 12.28 & Berkeley\,94 \\
TYC\,3986-1138-1 & 22$^{\rm h}$22$^{\rm m}$50$^{\rm s}_{\upf{.}}$4 & 55$^{\circ}$51$''$38$'$ & O6  & 12.14 & Berkeley\,94 \\
\hline
\end{tabular}
\begin{list}{Table Notes.}
\item Col.\,1: early type star name, Col.\,2: right ascencion (J2000),
Col.\,3: declination (J2000), Col.\,4: spectral type,
Col.\,5: absolute magnitude in $V$-band, Col.\,6: related cluster.
$^1$BD+55\,2722 is a binary with secondary spectral type unknown (ADS\,15834).
\end{list}
\end{table*}

For the known OCs (Teutsch\,127 and Berkeley\,94)
SIMBAD\footnote{\it http://simbad.u-strasbg.fr/simbad} lists some early type
stars (Table\,\ref{tab:ear}) with their positions, spectral types and $V$
magnitudes. We indicate the related clusters in the last column.

\begin{figure}
\resizebox{\hsize}{!}{\includegraphics{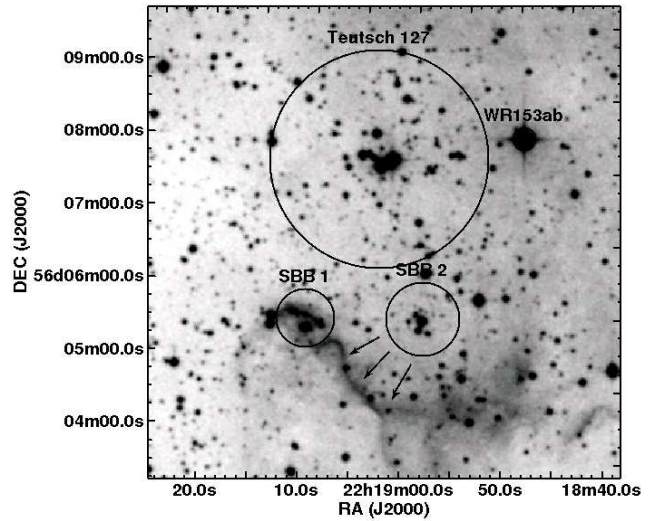}}
\caption[]{XDSS $R$-band image of the embedded star clusters in the center of
Sh2-132. The circles with radii listed in Table\,\ref{tab:clu} indicate
Teutsch\,127, SBB\,1 and SBB\,2. WR\,153ab is the brightest star to the
right of Teutsch\,127. Arrows indicate a possible bow-shock
below SBB\,2.}
\label{fig:t127}
\end{figure}

\begin{figure}
\resizebox{\hsize}{!}{\includegraphics{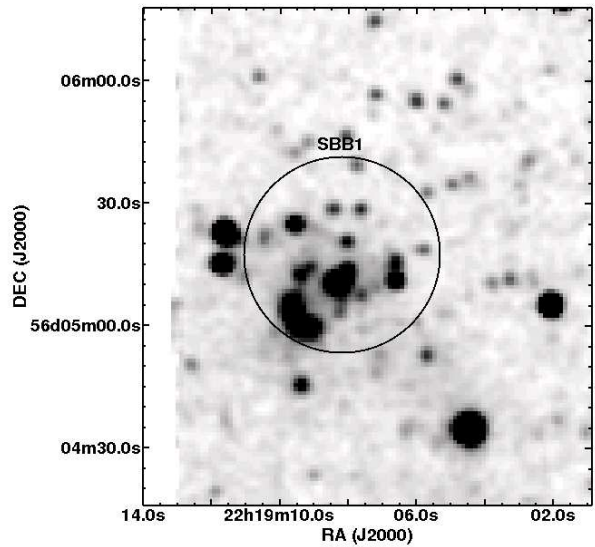}}
\caption[]{2MASS {\it K$_S$}-band image of the embedded infrared cluster
SBB\,1. The circle radius is given in Table\,\ref{tab:clu}.}
\label{fig:newcl1}
\end{figure}

\begin{figure}
\resizebox{\hsize}{!}{\includegraphics{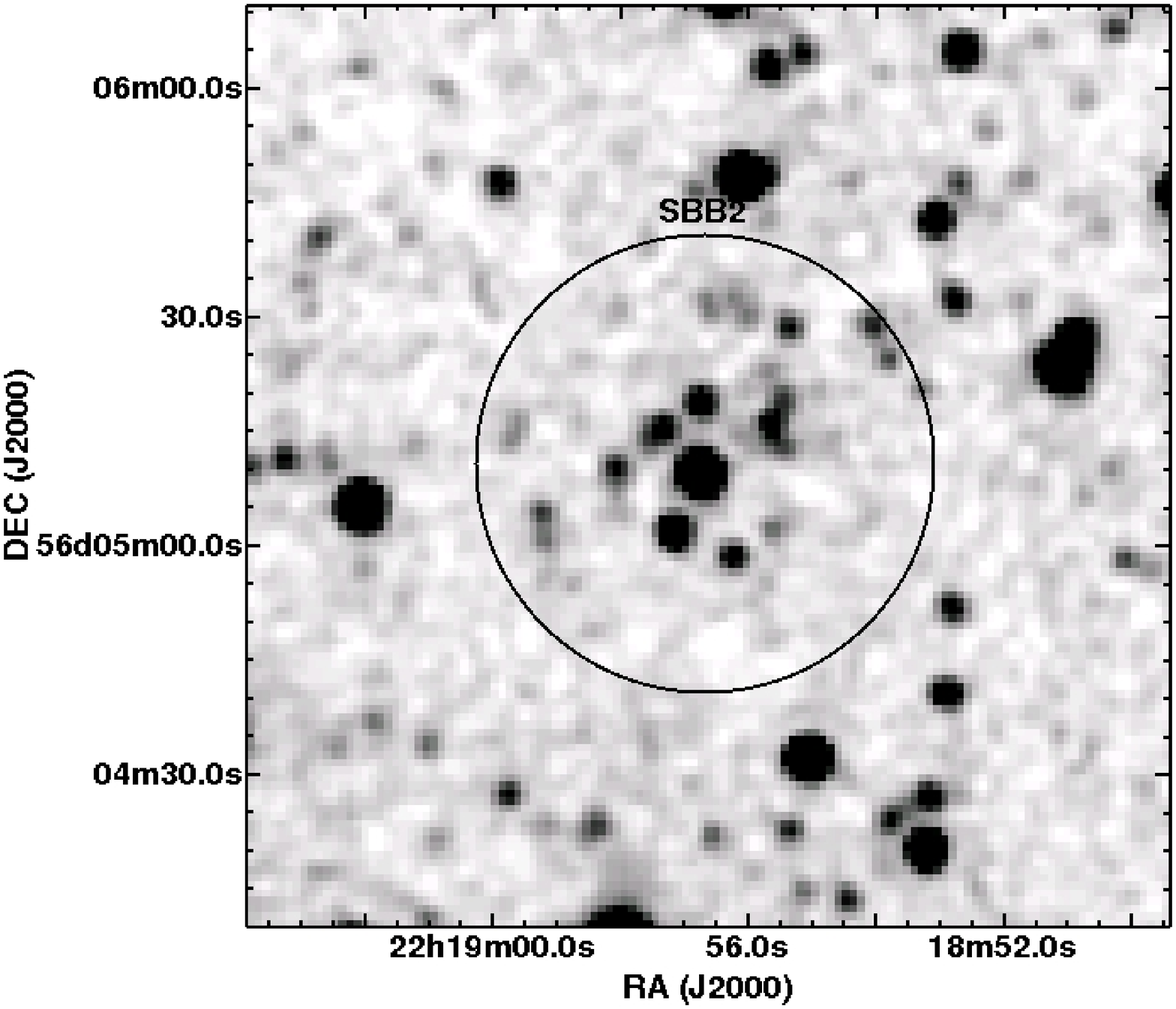}}
\caption[]{2MASS {\it K$_S$}-band images of the embedded optical cluster
SBB\,2. The circle radius is given in Table\,\ref{tab:clu}.}
\label{fig:newcl2}
\end{figure}

\begin{figure}
\resizebox{\hsize}{!}{\includegraphics{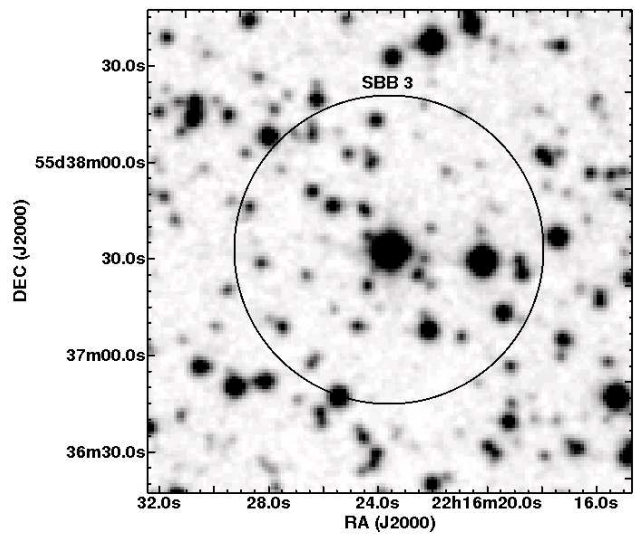}}
\caption[]{XDSS $R$-band image of the star cluster SBB\,3 surrounding WR\,152.
The circle radius is given in Table\,\ref{tab:clu}.}
\label{fig:wr152}
\end{figure}

\begin{figure}
\resizebox{\hsize}{!}{\includegraphics{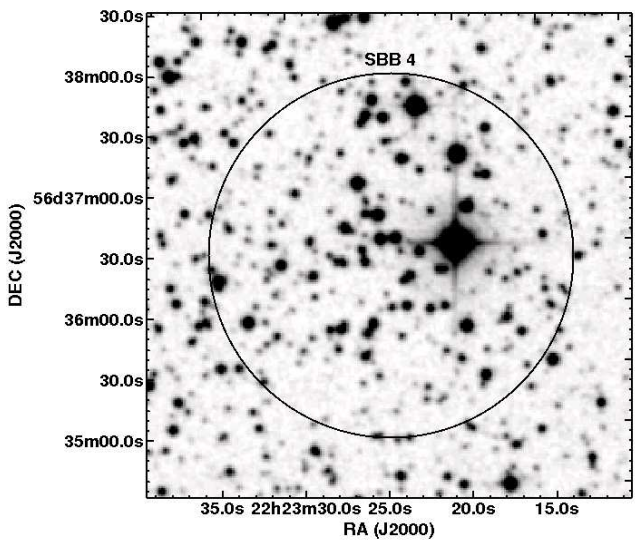}}
\caption[]{XDSS $R$-band image of the star cluster SBB\,4 in the outskirts of
Sh2-132. The circle radius is given in Table\,\ref{tab:clu}. The brightest star
is HD\,239944.}
\label{fig:newcl4}
\end{figure}

Fig.\,\ref{fig:t127} shows an XDSS $R$-band image containing the embedded star
clusters Teutsch\,127, SBB\,1, and SBB\,2, together with WR\,153ab star.
SBB\,1 and Teutsch\,127 are very similar in 2MASS bands. However, SBB\,1
becomes invisible in the $B$-band of DSS. This led us to refer to SBB\,1 and
Teutsch\,127 as `twins of different colours'. Figs.\,\ref{fig:newcl1} and
\ref{fig:newcl2} show blow-up 2MASS images in the {\it K$_S$}-band of the
embedded infrared cluster SBB\,1 and the embedded optical cluster SBB\,2,
respectively. Fig.\,\ref{fig:wr152} shows an XDSS $R$-band image of the faint
compact star cluster SBB\,3 surrounding the WR\,152 star.
Fig.\,\ref{fig:newcl4} shows an XDSS $R$-band image of the star cluster SBB\,4
in the outskirts of Sh2-132. Images of Berkeley\,94 can be found in WEBDA and
XDSS.

\section{CMD analysis}
\label{sec:cmd}
\subsection{CMD construction}
Photometry in the 2MASS $J$, $H$, and {\it K$_S$} bands was extracted in circular
fields of radius 40$'$ with
VizieR\footnote{\it http://vizier.u-strasbg.fr/viz-bin/VizieR?-source=II/246}
catalogue II/246,
centred on the coordinates listed in Table\,\ref{tab:clu}. As a quality
constraint, we only considered stars with photometric errors $\leq$\,0.1\,mag in
every bands.

CMDs displaying {\it J$\times$(J$-$H)} and {\it J$\times$(J$-$K$_S$)} for the
cluster stellar content and field stars for equal area (note that the
decontamination itself is based on larger fields to improve statistics) as the
clusters are shown in Figs.\,\ref{fig:t127_cmd} to \ref{fig:newcl4_cmd}.

Since proper motions are not available for a statistically significant sample of
stars in the target area, the field star decontamination approach used in this
work is based essentially on photometric properties. Given a spatial extraction
where a cluster is supposed to exist, and a comparison field (usually large
enough to provide statistical representativity in terms of colours and magnitude
-- e.g. \citealt{bonbic09b}), the method compares the density of stars, within
equivalent CMD cells, in both regions. To take full advantage of the 2MASS
photometry, the geometry of the CMD cells is three-dimensional, with axes along
$J$, {\it (J$-$H)} and {\it (J$-$K$_S$)}. The size step along each dimension
is ${\it \Delta} J$\,=\,1.0 and
${\it \Delta} (J$$-$$H)$\,=\,${\it \Delta} (J$$-$$K_S)$\,=\,0.25. The cluster
extraction radius is given in Table\,\ref{tab:clu} and the comparison field
range is {\it r}\,=\,20$'$$-$40$'$. Basically, for each cluster CMD cell we
compute an estimate of the field contamination based on the comparison field.
Photometric uncertainties are taken into account, in the sense that what we
compute is the probability of a given star to be found in a cell (this reduces
to the difference of the error function between the borders of the cells).
Summing the probability for all stars, the end result of this procedure is the
number-density of stars in a given cell. We do this for the cluster and
comparison field cells. Next, the comparison field number-density
is converted back into an integer number of stars, which is subtracted
from the cluster extraction, on a cell-by-cell basis.
Thus, the number of member stars in a given cell is $N^{cell}_{clean}$.
Allowing for variations in
the cell positioning corresponding to 1/3 of the adopted cell size in each
dimension, we are left with 27 different setups. Each setup produces a total
number of member stars, $N_{mem}\,=\,\sum_{cell}N^{cell}_{clean}$, from which we
compute the expected total number of member stars $\left<N_{mem}\right>$ by
averaging out $N_{mem}$ over all combinations. Stars are ranked according to the
number of times they survive all runs, and only the $\left<N_{mem}\right>$
highest ranked stars are considered cluster members and transposed to the
respective decontaminated CMD (bottom panels of Figs.\,\ref{fig:t127_cmd} to
\ref{fig:newcl4_cmd}). The difference between the expected number of field stars
(which may be fractional) and the number of stars effectively subtracted (which
is integer) from each cell is the subtraction efficiency, which summed over all
cells is 93.6$\pm$5.0\% for Teutsch\,127, 26.4$\pm$33.3\% for SBB\,2,
87.1$\pm$3.7\% for Berkeley\,94, 92.1$\pm$2.4\% for SBB\,3 and 90.9$\pm$7.5\%
for SBB\,4. No star has been removed within the area of SBB\,1.

All cluster {\it J$\times$(J$-$K$_S$)} CMDs of this sample were fitted
by eye with 1\,Myr Solar-metallicity (suitable for the Galactic disk in
general) Padova isochrones (\citealt{marigo08}). In addition, PMS isochrones by
\citet*{siess00} for the ages 0.2, 1, 2, 3, 5, 10, and 20\,Myr, were used when
necessary. Isochrone fits for each cluster (Table\,\ref{tab:pho}) correspond to
a distance from the Sun {\it d$_{\odot}$}\,=\,3.6\,kpc (that matches that of
\citealt{fosrou03}). This distance locates Sh2-132 in the Orion-Cygnus arm
(\citealt{momany08}). Only SBB\,4 appears to be closer and is
not part of the complex Sh2-132. Isochrone fit for the CMD of this cluster
yields a distance from the Sun {\it d$_{\odot}$}\,=\,2.27\,kpc.

\begin{table}
\caption[]{Fundamental parameters derived for the star clusters.}
\label{tab:pho}
\renewcommand{\tabcolsep}{0.4mm}
\renewcommand{\arraystretch}{1.25}
\begin{tabular}{lccccccc}
\hline
Name&{\it E(J$-$H)}&{\it E(J$-$K$_S$)}&{\it (m$-$M)J}&$A_V$&{\rm $t_0$}
&{\rm $t_{MS}$} &{\rm $t$}\\
& & & & &(Myr)&(Myr)&(Myr)\\
(1)&(2)&(3)&(4)&(5)&(6)&(7)&(8)\\
\hline
Teutsch\,127 & 0.26 & 0.41 & 13.50 & 0.25 & 10 & 1 & $\sim$\,5  \\
SBB\,1 & 0.80 & 1.25 & 14.99 & 0.78 & 3  & 1 & $\sim$\,1  \\
SBB\,2 & 0.35 & 0.55 & 13.75 & 0.34 & 5  & 1 & $\sim$\,2  \\
Berkeley\,94 & 0.22 & 0.34 & 13.39 & 0.22 & 10 & 1 & $\sim$\,5  \\
SBB\,3 & 0.30 & 0.47 & 13.61 & 0.29 & 20 & 1 & $\sim$\,15 \\
SBB\,4 & 0.35 & 0.55 & 12.75 & 0.34 & 10 & 1 & $\sim$\,5  \\
\hline
\end{tabular}
\begin{list}{Table Notes.}
\item Col.\,1: star cluster name, Col.\,2: colour excess {\it E(J$-$H)},
Col.\,3: colour excess {\it E(J$-$K$_S$)}, Col.\,4: observed distance modulus
assuming a distance from the Sun {\it d$_{\odot}$}\,=\,3.6\,kpc, except for
SBB\,4 for which we derived {\it d$_{\odot}$}\,=\,2.27\,kpc, Col.\,5:
$V$-band absorption, Col.\,6: age from the oldest PMS isochrone fit,
Col.\,7: age from the MS isochrone fit, Col.\,8: adopted age.
\end{list}
\end{table}

\subsection{CMD results}
\label{sec:res}
\begin{figure}
\resizebox{\hsize}{!}{\includegraphics{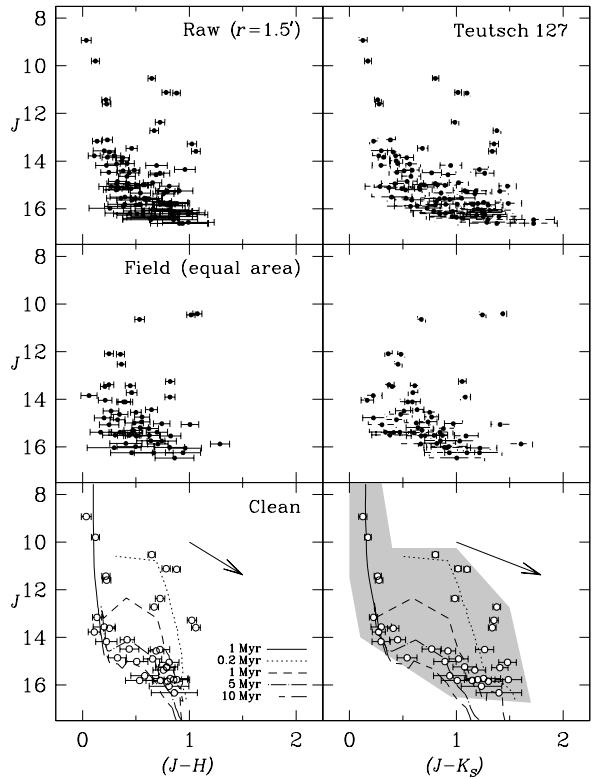}}
\caption[]{2MASS CMDs extracted from the {\it r}\,$<$\,1.5$'$ region of
Teutsch\,127. Top panels: observed photometry with
{\it J$\times$(J$-$H)} (left) and {\it J$\times$(J$-$K$_S$)} (right). 
Middle: equal-area (29.962$'$\,$<$\,{\it r}\,$<$\,30$'$) extraction from the
comparison field. Bottom panels: decontaminated CMDs fitted with 1\,Myr
Solar-metallicity Padova isochrone (\citealt{marigo08}) and
0.2, 1, 5 and 10\,Myr PMS isochrones
(\citealt{siess00}). The shaded area corresponds to the colour-magnitude filter.
The arrows indicate the reddening vectors for $A_V$\,=\,5.}
\label{fig:t127_cmd}
\end{figure}

\begin{figure}
\resizebox{\hsize}{!}{\includegraphics{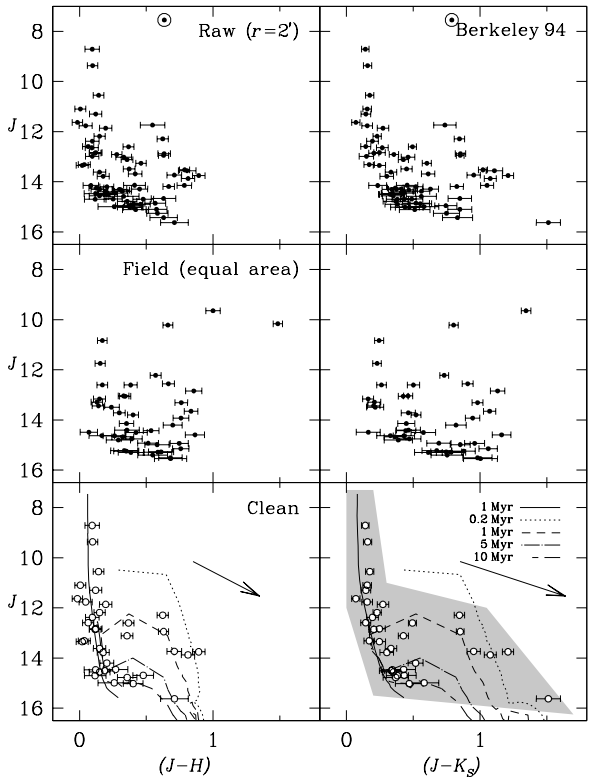}}
\caption[]{Same as Fig.\,\ref{fig:t127_cmd} for the region {\it r}\,$<$\,2$'$ of
Berkeley\,94 and comparison field between 29.933$'$\,$<$\,{\it r}\,$<$\,30$'$.
PMS isochrones are 0.2, 1, 5 and 10\,Myr. Circled points
represent the foreground TYC\,3986-341-1 star which is not part of the cluster,
althought has been assigned as member in SIMBAD.}
\label{fig:be94_cmd}
\end{figure}

\begin{figure}
\resizebox{\hsize}{!}{\includegraphics{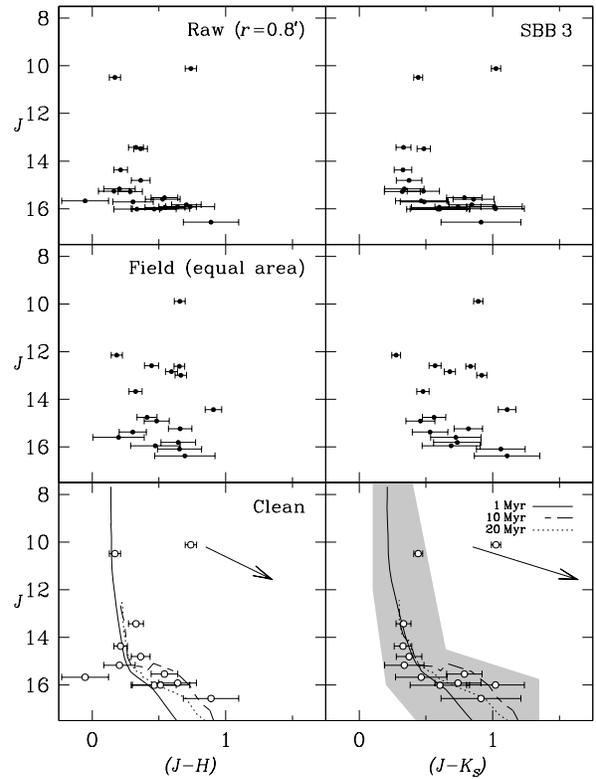}}
\caption[]{Same as Fig.\,\ref{fig:t127_cmd} for the region {\it r}\,$<$\,0.8$'$
of SBB\,3 and comparison field between 29.989$'$\,$<$\,{\it r}\,$<$\,30$'$.
PMS isochrones are 10 and 20\,Myr.}
\label{fig:wr152_cmd}
\end{figure}

\begin{figure}
\resizebox{\hsize}{!}{\includegraphics{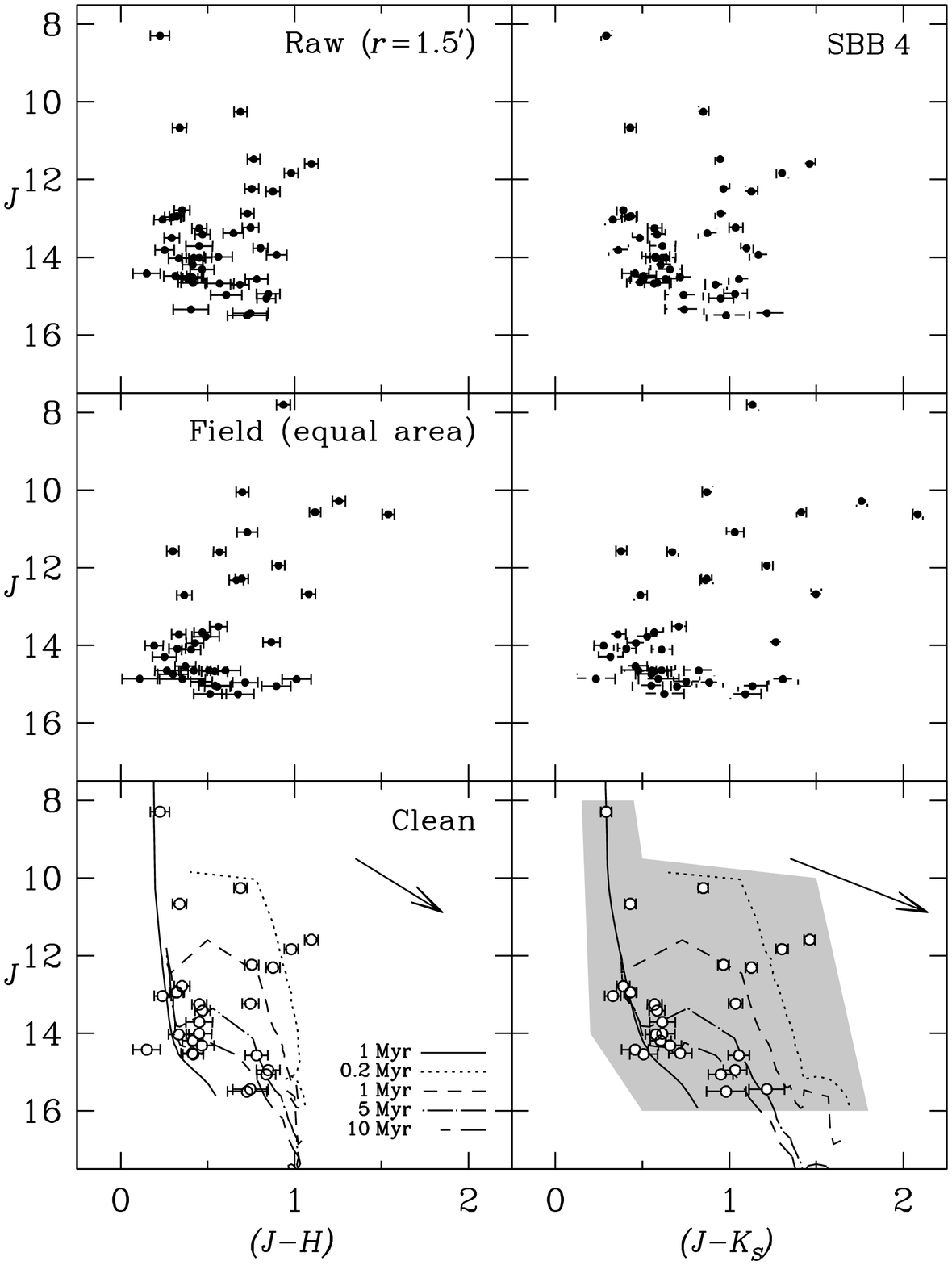}}
\caption[]{Same as Fig.\,\ref{fig:t127_cmd} for the region {\it r}\,$<$\,1.5$'$
of SBB\,4 and comparison field between 29.962$'$\,$<$\,{\it r}\,$<$\,30$'$.
PMS isochrones are 0.2, 1, 5 and 10\,Myr.}
\label{fig:newcl4_cmd}
\end{figure}

We detected MS and PMS stars in all clusters,
although Teutsch\,127 (Fig.\,\ref{fig:t127_cmd}) and Berkeley\,94
(Fig.\,\ref{fig:be94_cmd}) are more populated. In addition,
Teutsch\,127 may have a few Herbig Ae/Be stars heading to the MS.
These stars are located in the region {\it (J$-$K$_S$)}\,$>$\,0.75 and
$J$\,$<$\,12 (Fig.\,\ref{fig:t127_cmd}) in the CMD. The same occurs for
SBB\,4 (Fig.\,\ref{fig:newcl4_cmd}).

\begin{figure}
\resizebox{\hsize}{!}{\includegraphics{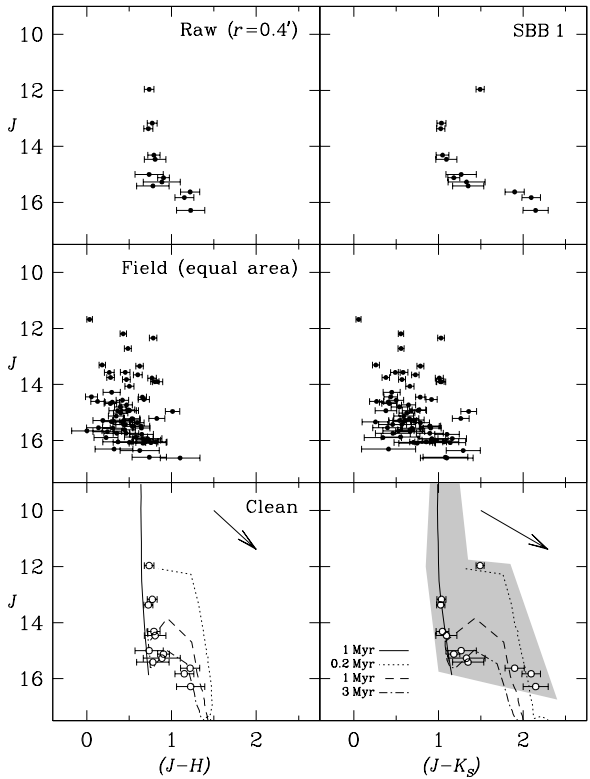}}
\caption[]{Same as Fig.\,\ref{fig:t127_cmd} for the region {\it r}\,$<$\,0.4$'$
of SBB\,1 and comparison field between 29.997$'$\,$<$\,{\it r}\,$<$\,30$'$.
PMS isochrones are 0.2, 1 and 3\,Myr.}
\label{fig:newcl1_cmd}
\end{figure}

\begin{figure}
\resizebox{\hsize}{!}{\includegraphics{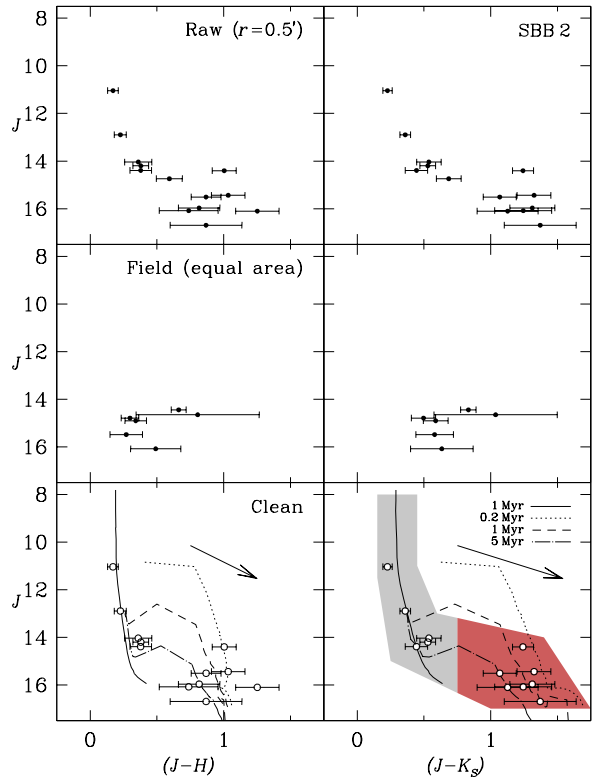}}
\caption[]{Same as Fig.\,\ref{fig:t127_cmd} for the region {\it r}\,$<$\,0.5$'$
of SBB\,2 and comparison field between 29.996$'$\,$<$\,{\it r}\,$<$\,30$'$.
PMS isochrones are 0.2, 1, and 5\,Myr.
Light-shadded polygon: colour-magnitude filter to isolate the MS stars.
Heavy-shadded polygon: colour-magnitude filter for the PMS stars.}
\label{fig:newcl2_cmd}
\end{figure}

The CMDs of Teutsch\,127 (Fig.\,\ref{fig:t127_cmd}), SBB\,1
(Fig.\,\ref{fig:newcl1_cmd}) and SBB\,2 (Fig.\,\ref{fig:newcl2_cmd}) show a
poorly populated MS, while the CMD of Berkeley\,94 (Fig.\,\ref{fig:be94_cmd})
has a better defined MS. Although each cluster shows PMS stars with ages
$<$\,10\,Myr, SBB\,3 has only lower-mass stars ({\it M}\,$<$\,2\,M$_{\odot}$)
with ages of 10$-$20\,Myr for the PMS. SBB\,2 has a detached MS and PMS,
therefore we filtered them separately (Fig.\,\ref{fig:newcl2_cmd}).

The star TYC\,3986-341-1 is present in SIMBAD as a member of Berkeley\,94 and
persists after CMD decontamination (Fig.\,\ref{fig:be94_cmd}), but its spectral
type is G8III, which is incompatible with shuch a young cluster. The distance
modulus derived from {\it B} and {\it V} values in SIMBAD reveals that it is
very close to the Sun and is not part of the star cluster.

\begin{figure}
\resizebox{\hsize}{!}{\includegraphics{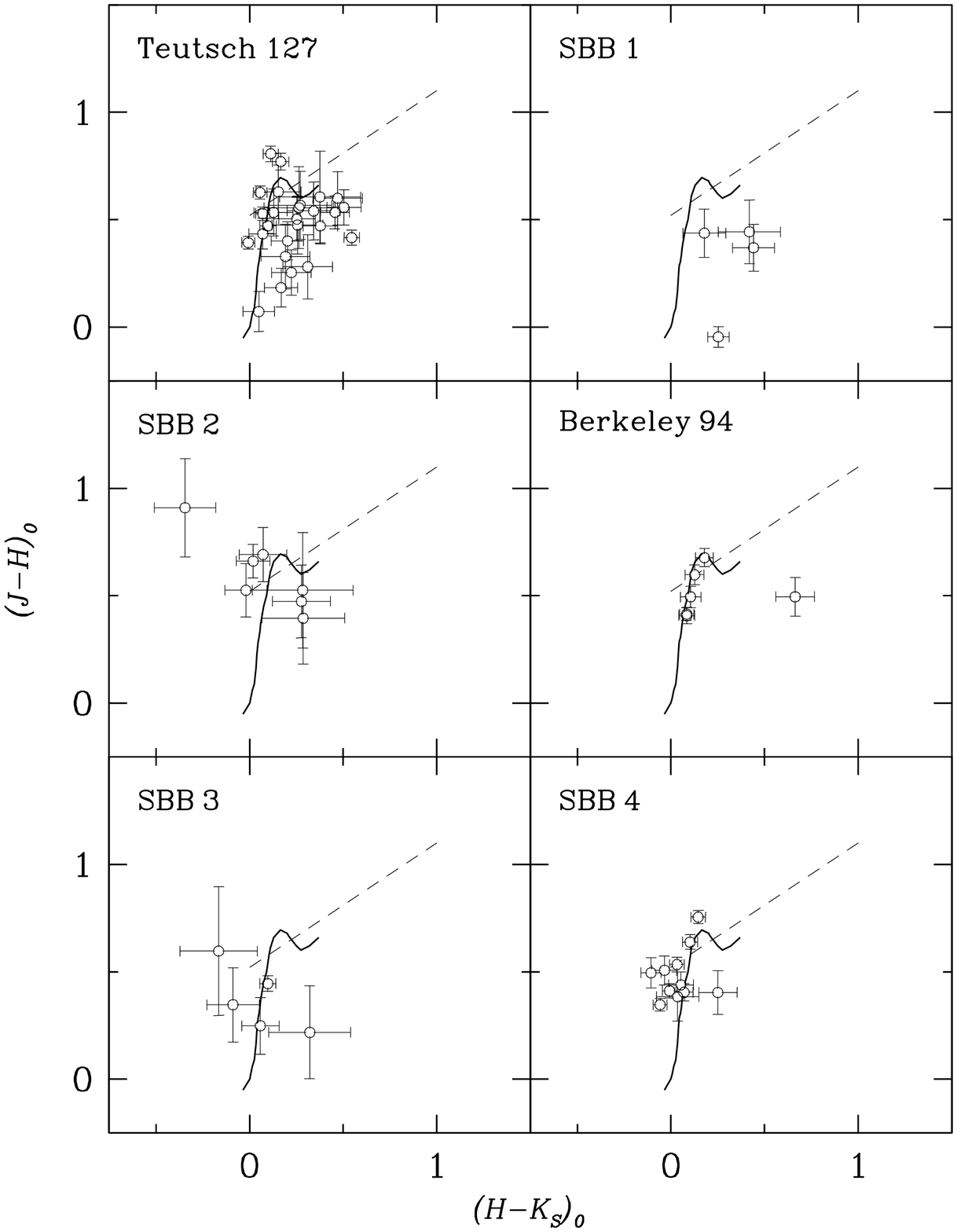}}
\caption[]{Absolute colour-colour diagrams {\it (H$-$Ks)$\times$(J$-$H)} of
PMS stars (open circles) of the clusters in the area of Sh2-132. Dashed line is
the CTTS locus (\citealt{meyer97}) and solid line is the dwarf star branch
(\citealt{besbre88}).}
\label{fig:ccor}
\end{figure}

Fig.\,\ref{fig:ccor} shows the intrinsic colour-colour diagrams
{\it (H$-$K$_S$)$\times$(J$-$H)} for the clusters in the area of Sh2-132.
Reddening values (Table\,\ref{tab:pho}) obtained by isochrone fits were
used to correct the magnitudes with the relations
{\it A$_J$/A$_V$}\,=\,2.76, {\it A$_H$/A$_V$}\,=\,0.176,
{\it A$_{K_S}$/A$_V$}\,=\,0.118, and {\it A$_J$}\,=\,2.76\,{\it E(J$-$H)} given
by \citet*{dsb02}, for {\it R$_V$}\,=\,3.1.
A comparison with the classical T Tauri star (CTTS) locus (dashed line) obtained
by \citet*{meyer97} shows that only a few of the PMS stars of our sample are T
Tauri stars. Another comparison of the colour-colour diagram with the standard
sequence for dwarf stars from \citet{besbre88} reveals the presence of some
dwarfs in all clusters, especially Berkeley\,94, Teutsch\,127 and SBB\,4.

The excess in {\it (H$-$K$_S$)} observed in the colour-colour diagrams for some
stars (Fig.\,\ref{fig:ccor}) suggests circumstellar disks, that characterise
many CTTSs (\citealt{furlan09}). These disks dissipate at $\sim$\,10\,Myr as
these stars evolve to weak-lined T Tauri stars and giant planet formation comes
to an end.

\begin{figure}
\resizebox{\hsize}{!}{\includegraphics{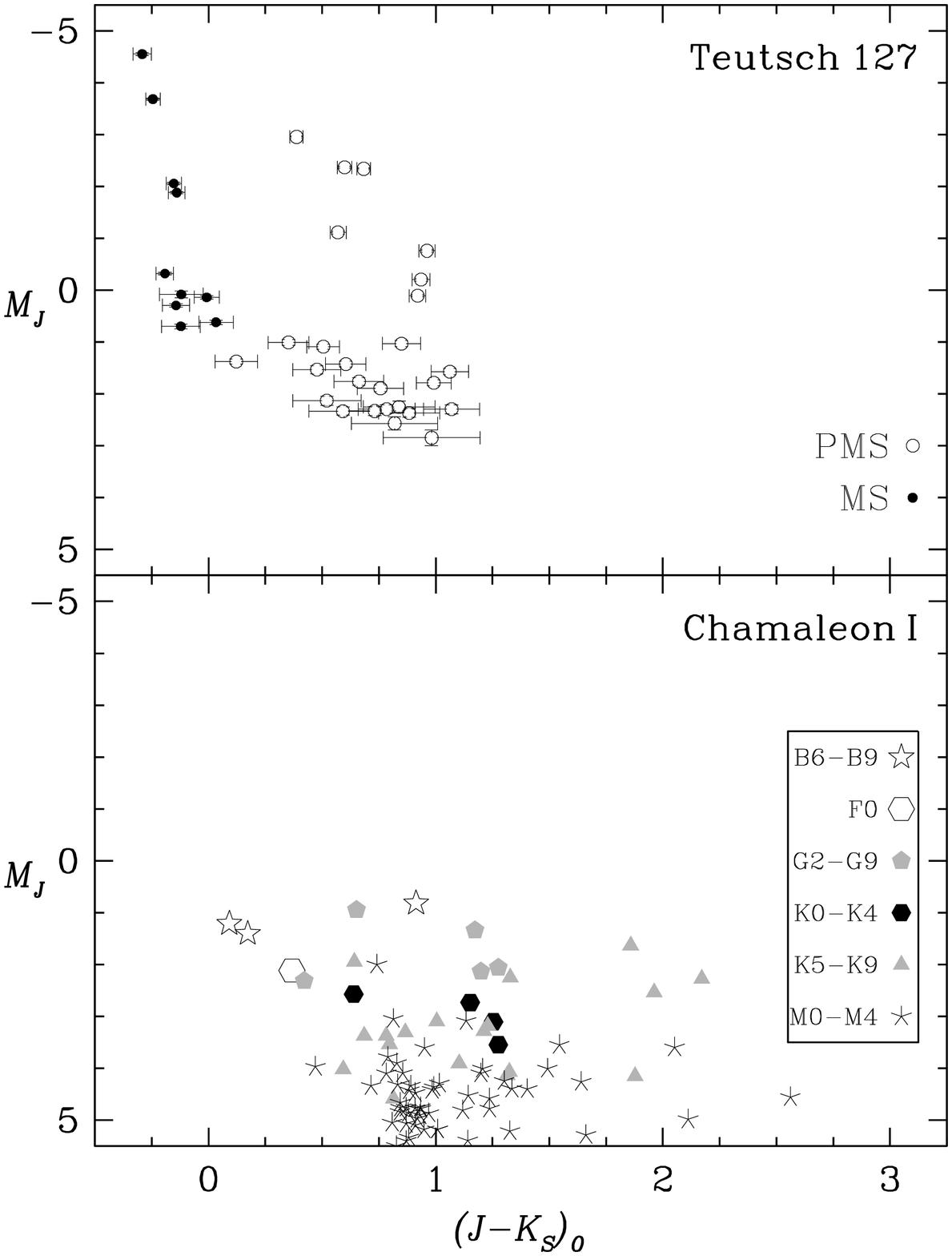}}
\caption[]{Comparison of the dereddened absolute CMD of Teutsch\,127
(top) with the dereddened absolute CMD of a PMS template (bottom) built with
data of the stellar content of Chamaleon\,I (\citealt{luhman07},
\citealt{luhman08}).}
\label{fig:temp}
\end{figure}

A comparison of the CMD of Teutsch\,127, taking into account absorption and
distances, with a PMS template (Fig.\,\ref{fig:temp}) built with
spectral types, magnitudes and colours of the stellar content of
Chamaleon\,I (\citealt{luhman07}, \citealt{luhman08} and references therein)
reveals that in Teutsch\,127, PMS stars appear to be distributed between
B6 (earliest) and M4 (latest), while in SBB\,2 they are bracketed by G5
and M4. In the other clusters, this distribution is more restrict. We identify
($i$) a B6 and a G5 in SBB\,1, ($ii$) F0, G8 and K3 types in SBB\,3, and ($iii$)
F0, G8, and K3 types in SBB\,4. The M4 lower limit is due to the observational
cutoff.

\subsection{CMD spread sources}
Open cluster CMDs can have a broader than usual MS because of several factors,
such as photometric errors, binarity, variability, field contamination, and the
most important in the case of embbeded clusters, differential reddening. These
effects can disturb isochrone fits. Photometric uncertainty is a function of
magnitude, so faint stars (such as PMS) have larger errors (\citealt{bonbic07}).

A large fraction of stars in clusters appears to be binary, so that points in
the CMD are shifted only above the isochrone, unlike other effects (photometric
errors and variability). \citet{platais07} found a binary fraction of 30\% for
IC\,2391 which has an age of 35\,Myr. \citet{cs2008} estimate in 76\% the binary
fraction in the Pleiades with an age of 125\,Myr. However, it is not yet clear
whether the reason for differences in binary fraction among clusters is age or
density. Furthermore, tidal stripping tears preferentially low-mass stars from
the edge of clusters, so that the binary fraction should increase, unless occurs
destruction of binaries.

Some PMS stars of these clusters may be T Tauri and therefore have
variable brightness causing spread of the points in the CMD. However, only
a few T Tauri stars are detectable with 2MASS in the present sample
(Sect.\,\ref{sec:res}) because of the large distance of Sh2-132
(Fig.\,\ref{fig:ccor}).

Differential reddening is the most important factor in the case of star clusters
located in regions with large quantities of gas and dust.
There is a remarkable large bubble centred close to the dark nebulae LDN\,1150 and
LDN\,1154 to the northeast of Sh2-132 (Fig.\,\ref{fig:sh132}). In fact,
reddening is larger in this direction according to the values
(Table\,\ref{tab:abs}) derived from the maps of dust infrared emission of
\citet*{sfd98}. To the west of Teutsch\,127 is the dark nebula LDN\,1161
(Fig.\,\ref{fig:sh132}).

These factors act cumulatively on the observed stellar magnitudes
scattering the points in the CMDs, but the PMS age spread appears to be a
major source of dispersion
(Figs.\,\ref{fig:t127_cmd} to \ref{fig:newcl4_cmd}).

\begin{table}
\caption[]{Absorptions derived from the maps of dust infrared emission of \citet{sfd98}.}
\label{tab:abs}
\renewcommand{\tabcolsep}{5.35mm}
\renewcommand{\arraystretch}{1.25}
\begin{tabular}{lcrr}
\hline
\multicolumn{1}{c}{Object} &\multicolumn{1}{c}{\it E(B$-$V)} &\multicolumn{1}{c}{\it A$_V$} &\multicolumn{1}{c}{\it A$_J$}\\
\multicolumn{1}{c}{(1)} &\multicolumn{1}{c}{(2)} &\multicolumn{1}{c}{(3)} &\multicolumn{1}{c}{(4)}\\
\hline
Teutsch\,127 & 5.888 & 19.520 & 5.311 \\
SBB\,1 & 5.361 & 17.773 & 4.836 \\
SBB\,2 & 6.100 & 20.220 & 5.502 \\
Berkeley\,94 & 0.734 & 2.433 & 0.662 \\
SBB\,3 & 1.399 & 4.638 & 1.262 \\
SBB\,4 & 0.998 & 3.310 & 0.901 \\
LDN\,1150 & 1.283 & 4.252 & 1.157 \\
LDN\,1154 & 2.143 & 7.104 & 1.933 \\
LDN\,1161 & 3.827 & 12.686 & 3.452 \\
\hline
\end{tabular}
\begin{list}{Table Notes.}
\item Col.\,1: object name, Col.\,2: reddening,
Col.\,3: {\it V}-band absorption, Col.\,4: {\it J}-band absorption.
\end{list}
\end{table}

\section{Stellar density profiles}
\label{sec:rdp}
In principle, CMDs of bound and un-bound young star clusters are
indistinguishable. Structural differences resulting of early changes in the
gravitational potential can show up in the stellar RDPs. We may expect that
young clusters with stellar dispersion have irregular RPDs that cannot be fitted
by any model based on an isothermal sphere (e.g. \citealt{kin62},
\citealt{wil75}, and \citealt*{eff87}).

\begin{figure}
\resizebox{\hsize}{!}{\includegraphics{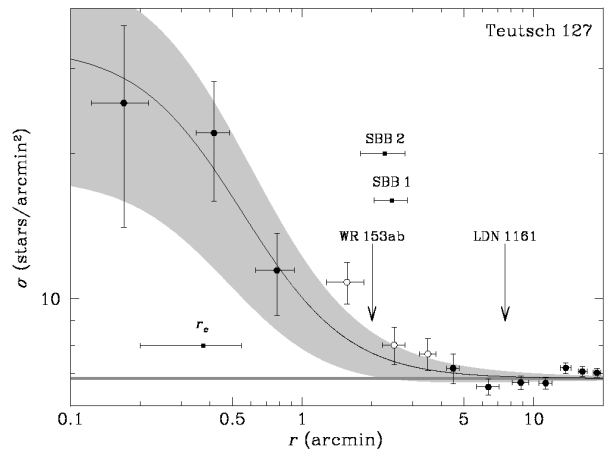}}
\caption[]{Stellar RDP for Teutsch\,127 built with colour-magnitude filter
(Fig.\,\ref{fig:t127_cmd}). The best-fit King like profile is shown as a solid
line. Open circles were excluded from the fit. Background stellar
level is represented by the shaded stripe in the bottom and ${\it 1\sigma}$ fit
uncertainty is represented by the shaded region along the fit. The core radius
($r_c$), WR\,153ab, SBB\,1, SBB\,2 and LDN\,1161 positions are indicated.}
\label{fig:t127_rdp}
\end{figure}

\begin{figure}
\resizebox{\hsize}{!}{\includegraphics{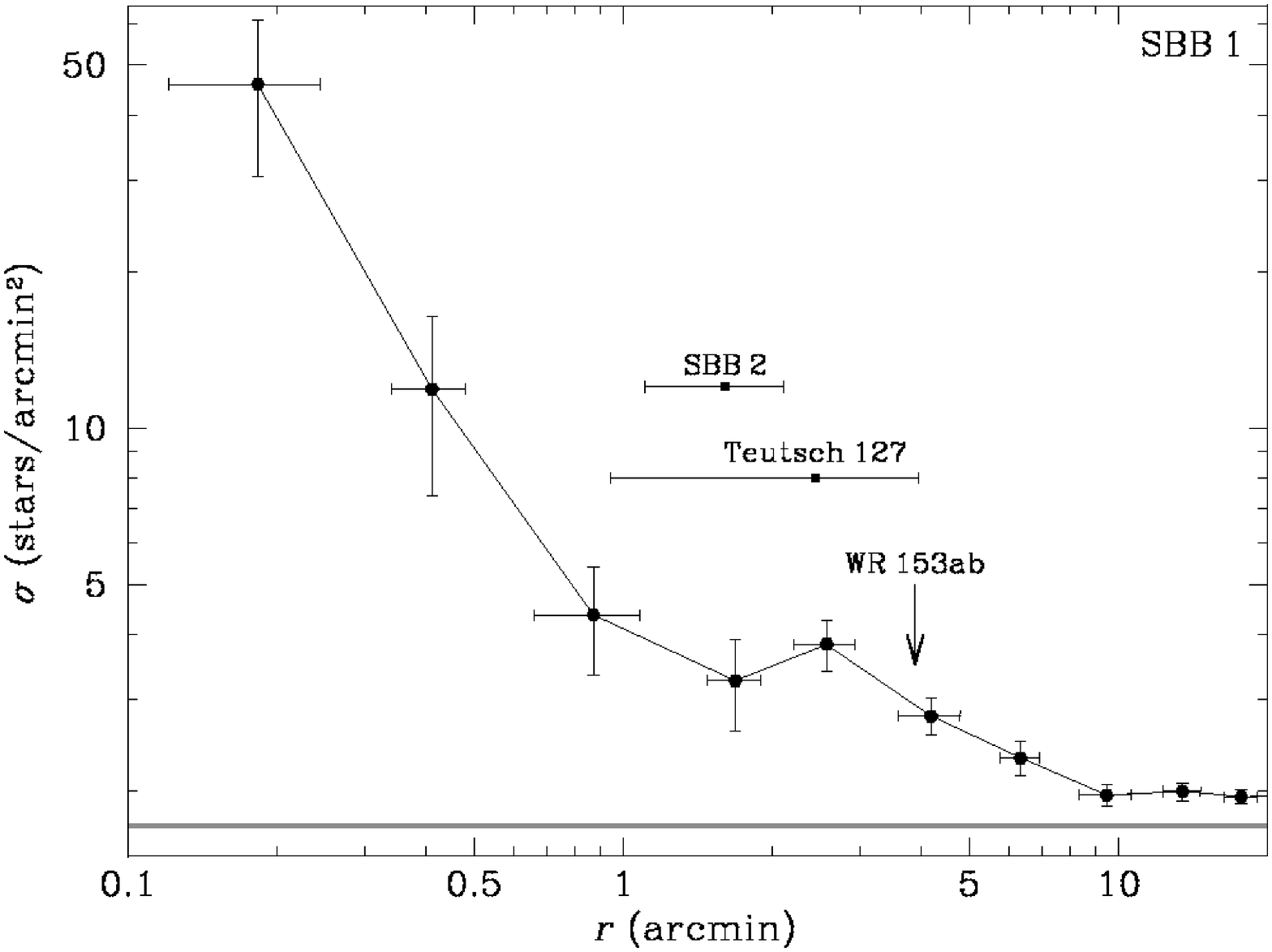}}
\caption[]{Stellar RDP for SBB\,1 built with colour-magnitude
photometry. WR\,153ab, SBB\,2 and Teutsch\,127 positions are indicated.}
\label{fig:newcl1_rdp}
\end{figure}

\begin{figure}
\resizebox{\hsize}{!}{\includegraphics{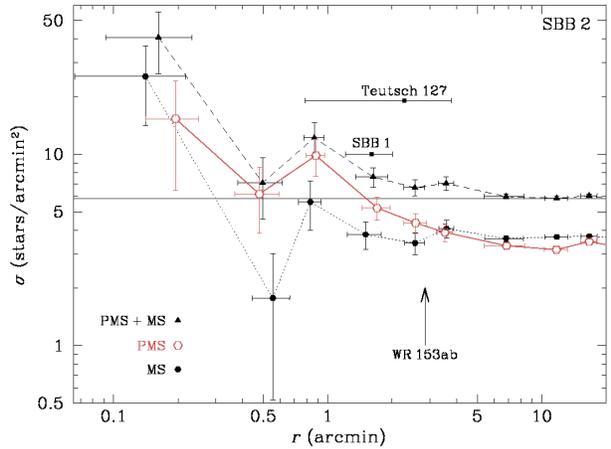}}
\caption[]{Stellar RDPs for SBB\,2 built separately for PMS stars (open
circles) and MS (filled circles) stars with colour-magnitude photometry.
Stellar RDP of PMS\,+\,MS is represented by filled triangles.
WR\,153ab, SBB\,1 and Teutsch\,127 positions are indicated.}
\label{fig:newcl2_rdp}
\end{figure}

\begin{figure}
\resizebox{\hsize}{!}{\includegraphics{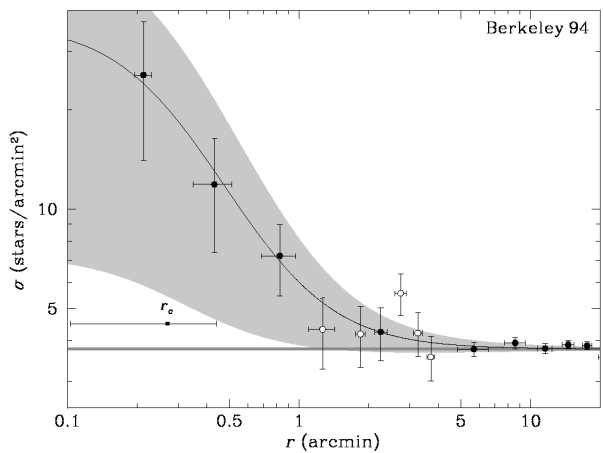}}
\caption[]{Same as Fig.\,\ref{fig:t127_rdp} for Berkeley\,94.}
\label{fig:be94_rdp}
\end{figure}

\begin{figure}
\resizebox{\hsize}{!}{\includegraphics{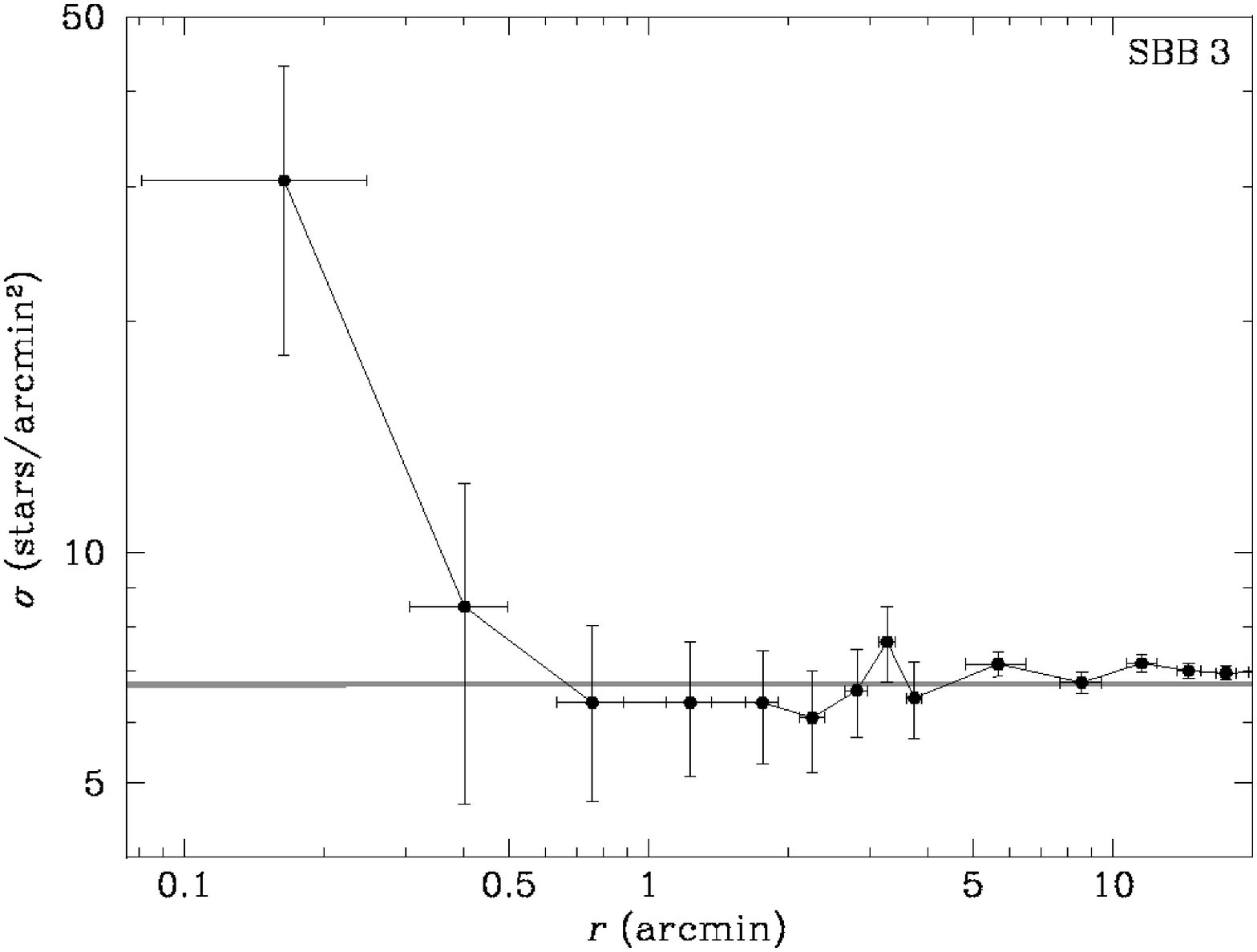}}
\caption[]{Same as Fig.\,\ref{fig:newcl1_rdp} for SBB\,3.}
\label{fig:wr152_rdp}
\end{figure}

\begin{figure}
\resizebox{\hsize}{!}{\includegraphics{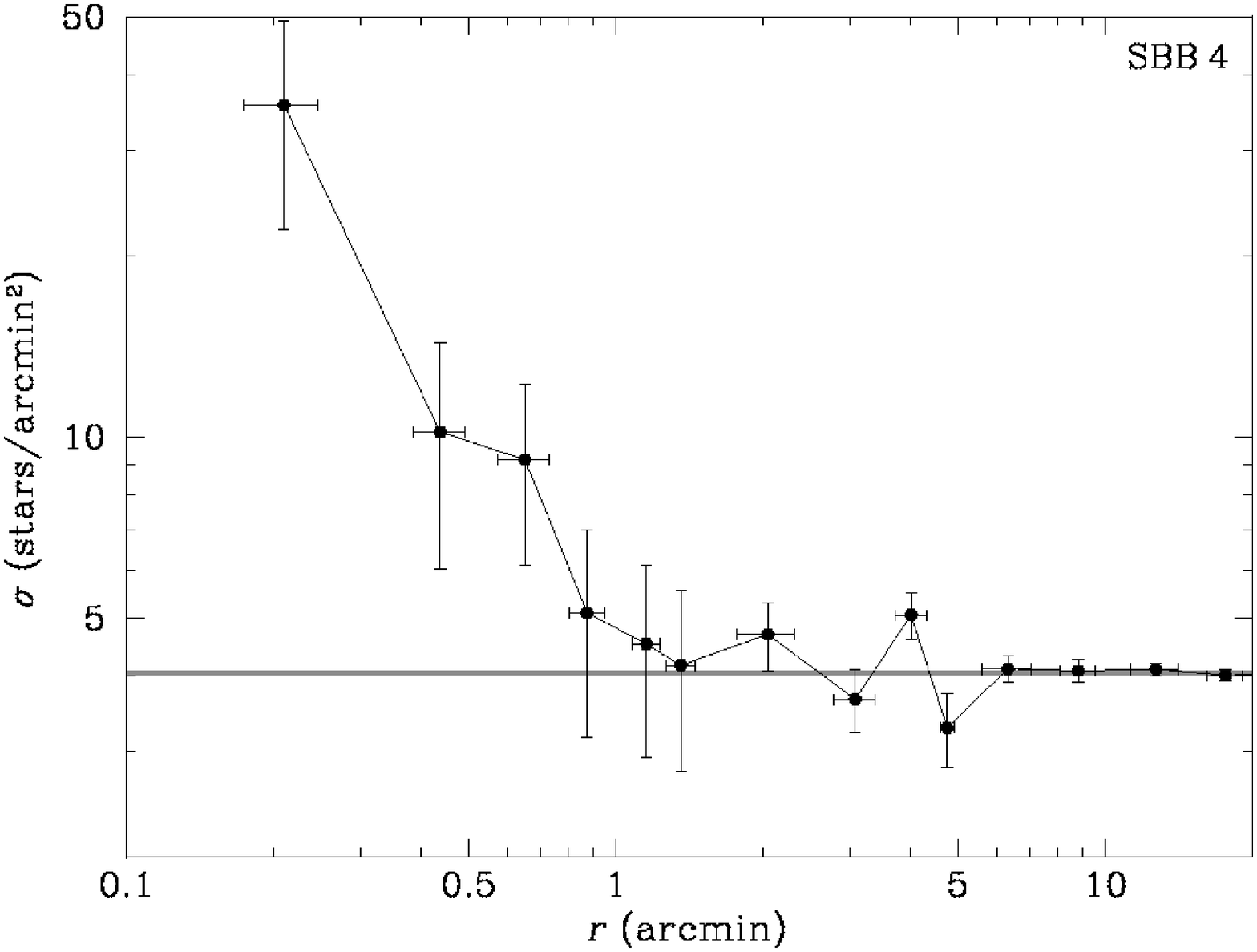}}
\caption[]{Same as Fig.\,\ref{fig:newcl1_rdp} for SBB\,4.}
\label{fig:newcl4_rdp}
\end{figure}

RDPs were built for the colour-magnitude filtered photometry
(Figs.\,\ref{fig:t127_rdp} to \ref{fig:newcl4_rdp}). Background densities
(${\it \sigma_{bg}}$) were measured in the field ({\it r}\,=\,20$'$$-$40$'$).
King-like profiles were fitted for the RDPs of Teutsch\,127 and Berkeley\,94
with a non-linear least-squares routine that uses errors as weights. The
structural parameters central surface density (${\it \sigma_0}$) and core radius
($r_c$) were obtained from this fit, while ${\it \sigma_{bg}}$ was kept
constant. The results are shown in Table\,\ref{tab:par}. The RDPs of SBB\,1,
SBB\,2, SBB\,3 and SBB\,4 could not be fitted by a King-like profile.

The presence of the dark nebula LDN\,1161 and its obscuring effect is clearly
observed as a depression in the RDP of Teutsch\,127 (Fig.\,\ref{fig:t127_rdp})
for 6$'$\,$<$\,{\it r}\,$<$\,11$'$.

The RDPs of central clusters (Teutsch\,127, SBB\,1 and SBB\,2) overlap for
radii larger than 1$'$. They might represent radial profiles of merging clusters
(\citealt{carva08}).

SBB\,2's RDP was divided into two components (Fig.\,\ref{fig:newcl2_rdp}),
one for MS and other for PMS stars. The total RDP (MS\,+\,PMS) is also shown.
The higher central density of MS stars
may suggest primordial mass segregation with an
outer region dominated by lower-mass PMS stars.

The RDP of SBB\,3 (Fig.\,\ref{fig:wr152_rdp}) results in a dense and compact
core possibly sustained by the gravitational potential of WR\,152.

Finally, SBB\,4 (Fig.\,\ref{fig:newcl4_rdp}) has an irregular RDP with some
bumps which can be evidence of stellar dispersion.

\begin{table}
\caption[]{Structural parameters of the star clusters.}
\renewcommand{\tabcolsep}{1.3mm}
\renewcommand{\arraystretch}{1.25}
\label{tab:par}
\begin{tabular}{lcccc}
\hline
Name&{${\it \sigma_0}$}&$r_c$ &$r_c$&{${\it \sigma_{bg}}$}\\
&(stars/\arcmin$^{\rm 2}$)&(\arcmin)&(pc)&(stars/\arcmin$^{\rm 2}$)\\
(1)&(2)&(3)&(4)&(5)\\
\hline
Teutsch\,127 & 27.05$\pm$19.38 & 0.37$\pm$0.21 & 0.39$\pm$0.22 & 6.39$\pm$0.04 \\
SBB\,1 & -- & -- & -- & 1.71$\pm$0.02 \\
SBB\,2 & -- & -- & -- & 5.87$\pm$0.04 \\
Berkeley\,94 & 32.72$\pm$29.23 & 0.27$\pm$0.17 & 0.28$\pm$0.18 & 3.77$\pm$0.03 \\
SBB\,3 & -- & -- & -- & 7.94$\pm$0.05 \\
SBB\,4 & -- & -- & -- & 4.05$\pm$0.03 \\
\hline
\end{tabular}
\begin{list}{Table Notes.}
\item Col.\,1: star cluster name, Col.\,2: central density,
Col.\,3: core radius (arcmin), Col.\,4: core radius (pc) assuming
a distance from the Sun {\it d$_{\odot}$}\,=\,3.6\,kpc,
Col.\,5: background density.
\end{list}
\end{table}

\section{Discussion}
\label{sec:dis}
The morphology of Sh2-132, which contains several bubbles suggests
triggered star formation similarly to Sh2-284 (\citealt{puga09}). Additional
small star clusters not yet identified are possible in the area of the Sh2-132
complex. A higher resolution mid-infrared study would be important.

Teutsch\,127 appears to be a young OC with age $\sim$\,5\,Myr, while SBB\,1
seems to be $\sim$\,1\,Myr old. SBB\,1 is deeply embedded and
may result from a stellar feedback of O and B-type stars in somewhat older
surrounding clusters. It is superimposed on a probable
bow-shock $\sim$\,3$'$ to the southwest of Teutsch\,127 and WR\,153ab.
However, this bow-shock seems to have been generated in
SBB\,2. We may be witnessing sequential effects of star formation.

The presence of the trapezium system Trap\,900 in the center of Teutsch\,127
will certainly have an important role in the cluster evolution. Trapezium
systems evolve into hierarchical systems (with a much larger separation among
its components) or even disperse in a few million years producing runaway
stars. The oldest one identified so far has $\sim$\,50\,Myr
(\citealt{abtcor00}).

Teutsch\,127 is close to WR\,153ab that is a spectroscopic binary with a primary
WN6o and an O6I (\citealt*{ssm96}). Evolution of this WR to spectral type WC and
its subsequent explosion as supernova will certainly have an impact on the
dynamical evolution of neighbouring clusters (Teutsch\,127, SBB\,1 and
SBB\,2), by means of gas removal.

SBB\,1 is the youngest cluster in the sample and remains embedded. The
residual gas expulsion and stellar evolution may cause an increase of core
radius.

Since SBB\,2 expulsed its residual gas, star formation must have stopped and
its survival as a bound OC depends on the dynamics and evolution of neighbouring
clusters and specially of the WR\,153ab star.

SBB\,3 hosts the WR\,152 star of type WN3(h)-w, a mass of
12\,M$_{\odot}$ and mass loss rate
$\log \dot{M}$(M$_{\odot}$/yr)\,=\,-5.5 (\citealt*{hgl06}).
Since WR\,152 has a relatively low mass for a WR star (the minimum initial mass
for a star to become a WR at Solar-metallicity is 25\,M$_{\odot}$) it can
explode as supernova without evolving through the WC type, expulsing gas and
dissolving the star cluster very early.

Two possible destinations are suggested for the central clusters of Sh2-132:

\begin{enumerate}
\item Given the proximity among these clusters, a merger may
occur, which would generate a more massive OC. As a consequence this cluster
would appear dynamically older than the evolution of its stars indicates, that
is, with mass segregation more advanced than expected (\citealt{moebon09}).
\item Alternatively, the dissolution of the clusters, as a
consequence of residual gas expulsion, stellar evolution and dynamics of the
trapezium system. In this case, Trap\,900 might leave a compact fossil remnant.
\end{enumerate}

\section{Concluding remarks}
\label{sec:con}
We report the discovery of 4 star clusters (here referred to as SBB\,1,
SBB\,2, SBB\,3 and SBB\,4) in the area of the Sh2-132 complex. These
clusters and two previously catalogued ones (Teutsch\,127 and Berkeley\,94) were
analyzed with 2MASS photometry and show evidence that the Sh2-132 complex
is a site of triggered star formation, as suggested by its hierarchical
structure and age distribution of the star clusters. The presence of gas
and dust bubbles reinforces this idea.

\section*{Acknowledgments}
We thank an anonymous referee for constructive
comments and suggestions.
We acknowledge support from the Brazilian Institution
CNPq. This publication makes use of data products from
the Two Micron All Sky Survey, which is a joint project of
the University of Massachusetts and the Infrared Processing
and Analysis Centre/California Institute of Technology,
funded by the National Aeronautics and Space Administration
and the National Science Foundation. This research has
made use of the WEBDA database, operated at the Institute
for Astronomy of the University of Vienna.

\label{lastpage}

\end{document}